# nanoHUB.org: Experiences and Challenges in Software Sustainability for a Large Scientific Community

Lynn Zentner, Michael Zentner, Victoria Farnsworth,
Michael McLennan, Krishna Madhavan, and Gerhard Klimeck
*Network for Computational Nanotechnology, Purdue University*

## Background

The Network for Computational Nanotechnology Cyberinfrastructure Node (NCN CI) is responsible for the operation and support of the cyber-platform nanoHUB.org, serving a large scientific community centered around the nanoscience and nanotechnology fields. The word "community" is constantly on our minds. One definition of community is a "body of persons of common and especially professional interests scattered through a larger society.[1]" For us, this means we have a challenge of supporting a variety of community members, with varying viewpoints and needs; generally many wish to participate and utilize nanoHUB while a significantly smaller subset contribute to our online facility for that community. The staunchness of policy we set, the degree to which we enforce it, and the features we choose to make available are under constant examination: Are we making decisions that selectively encourage some members of the community while alienating others? Is the community we serve only those who agree with the policies we set, or do we need to look at setting policies that do not unnecessarily exclude well-respected members of the nanoscience and nanotechnology fields? nanoHUB has a significant ten year history of data and experience. nanoHUB users exceed 260,000 annually and access a portfolio of over 4000 resources contributed by over 1,000 authors, including over 300 simulation tools contributed by 411 software authors and developers. Such broad use and a large, vibrant community provide continued opportunities for growth, but careful management of policies and processes is necessary to anticipate and meet associated challenges.

## The nanoHUB Community

The nanoHUB community can be thought of as being composed of several overlapping groups of stakeholders. At its core, nanoHUB serves its user community with cutting edge tools and learning materials that may be so new as to not yet have been presented in textbooks. This user community consists of users in both the research and educational arenas and benefits from and helps speed the transition of research code into use by other researchers as well as into the classroom. We have documented the rapid transition of research codes into the classroom, on average, in less than 6 months from the initial publication of the code on nanoHUB.

A second group served by nanoHUB is the group of researchers who themselves are developing code related to their scientific research areas. nanoHUB, through its easy to use Rappture Development Toolkit[2], allows the scientists, with minimal training, to deploy and maintain their code in a way that is easily accessible. This approach eliminates the "middle man," a computer scientist previously required to rebuild and maintain code for use on the web, effectively disenfranchising the original author. nanoHUB presents the originating scientists with an opportunity to share their research products easily and continue to stay involved in the ongoing support, maintenance, and enhancement of their code, with significant and measureable impact.

Lastly, nanoHUB plays an active role in the cyberinfrastructure community. nanoHUB found such success within its own scientific community that the infrastructure powering it was extracted in order to bring similar HUB technology to other scientific areas[3]. The result is that nanoHUB now contributes to and benefits from development efforts to expand and improve the functionality of the core infrastructure, known as HUBzero. The nanoHUB cyberinfrastructure does not operate in a vacuum, but rather takes the opportunity to leverage and incorporate appropriate technologies, such as Pegasus workflow management tools[4], to the benefit of the nanoHUB users and tool developers.

## Opportunities and Challenges

Managing and growing a successful cyberinfrastructure such as nanoHUB presents a variety of opportunities and challenges, particularly in regard to software. nanoHUB is in the somewhat unique position of dealing with issues related to two types of software: the open source HUBzero software that powers the infrastructure as well as the many scientific codes contributed and deployed by the nanotechnology community. Over the years, nanoHUB has explored several issues related to software deployment and publishing, including licensing, intellectual property, export control, incentives, and quality.

*Maximizing Participation Relies on Tolerant Licensing*

nanoHUB has repeatedly considered the implications of licensing, both with regard to its core HUBzero platform software and with respect to scientific code contributed by its community members. The HUBzero code[5] is available through regular open source release under the LGPLv3 license. This licensing allows independent developers and HUB owners to create their own unique components within the HUBzero framework and license those works as they choose. If they make changes to the source code and redistribute the derivative work to anyone else, they are required to keep the same license on their code and post it publicly. Developers are encouraged to feed changes back to HUBzero, such that they can be considered for the next open source release. This license and approach was selected in order to encourage open collaboration without being overly restrictive.

A similar context drives how we handle licensing of scientific codes on nanoHUB. We believe that a steady stream of high quality, open access content is necessary to continue to grow and maintain a vibrant community and strive to lower barriers to dissemination of content on nanoHUB.

When an author submits code on nanoHUB, we provide as much flexibility as possible to them so that they can contribute code to the community and still meet the requirements of their funding agency, institution, as well as their personal intellectual property concerns. In general all the codes need to be in an *open access* form, where any nanoHUB user can run and execute the scientific codes via a graphical user interface, much like an app on an iPhone, except that the science codes on nanoHUB run in the cloud on behalf of the user. We provide an opportunity for contributors to license their code as *open source*, and have found that out of 323 currently published tools, only 16 tools have utilized an open source license, with their authors choosing a variety of flavors of that license, ranging across GPL, NCSA, BSD, and LGPL. While scientists can and will share their tools with the community through the nanoHUB open access policy, the above numbers indicate that a *requirement* by a cyberinfrastructure such as nanoHUB for contributors to share their source code through an open source license would drastically and negatively affect the sharing of tools that we have seen historically. *We believe that the cyberinfracture can play a role in providing education on the benefits and best practices of open source release, but ultimate choice of the actual license belongs to the tool authors, funding agencies, and supporting institutions.*

### *World Complexities Demand Flexible Software Access*

Another issue nanoHUB has needed to consider is related to restrictions regarding export control. The content authors carry the responsibility of knowing whether there are any export restrictions on the code they deploy on nanoHUB. Our contribution process allows them to restrict access to their code accordingly, allowing the choice of full access, restriction to US users only, restriction to non-D1 nations, or in the case of commercial software that may be licensed only to a particular set of users, licensing to particular groups. We maintain that the above choices allow us to support the greatest number of users with the greatest number of tools, which allowing contributors to control access in a way that meets any institutional, funding agency, or commercial requirements.

### *Incentives and Low Barriers to Participation Keep the Content Pipeline Flowing*

A last set of considerations revolves around incentivizing contributions while maintaining quality. As mentioned above, we strive to lower barriers to contribution in order to maintain a steady flow of content to our community. However, we must balance the ease of contribution with maintaining a level of quality in our contributed software. Software contributors are strongly encouraged to provide at least minimal documentation, such as a first time user guide as well as scholarly publications that support the scientific approach and content of the code. nanoHUB also provides a mechanism for developers to create regression testing suites for their code, such that code revisions can be vetted against these tests.

With over ten years of experience in hosting scientific tools, the nanoHUB team has concluded that the user community can be a strong partner in crowd-sourcing quality control. Through open, transparent mechanisms such as reviews, question and answer forums, wishlists, citation counts, and usage statistics, it is easy for users to see which tools are actively used and maintained, and the highest quality tools are allowed to bubble to the top. These same mechanisms provide an incentive for authors to contribute and maintain their codes, providing both a heartbeat of the quality and usefulness of a particular tool as well as quantifiable measure of the tool and author's impact on the scientific community. See for example Figure 1 which shows the snapshot view of the very popular tool *"Bandstructure Lab"* on nanoHUB.org. This[6] and any other published tool has a digital object identifier that can be cited in scientific publications. Lastly we and (most of) the tool authors view the tools as a publication which carries the author's name publicly. It is therefore in the responsible author's interest to deliver high quality material.

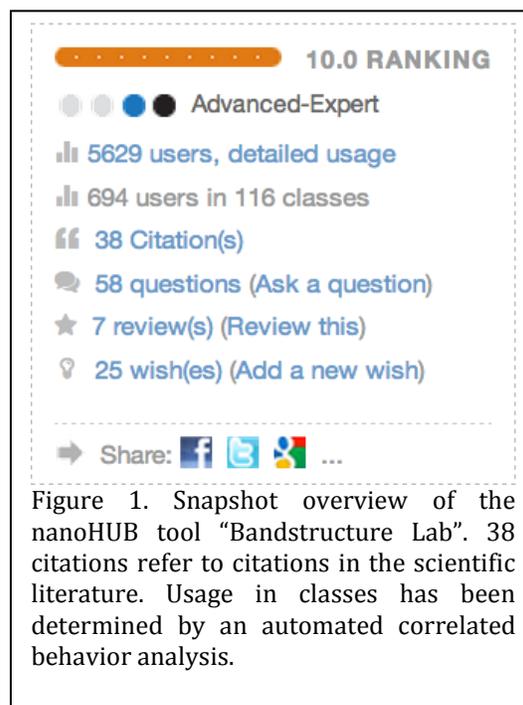

Figure 1. Snapshot overview of the nanoHUB tool "Bandstructure Lab". 38 citations refer to citations in the scientific literature. Usage in classes has been determined by an automated correlated behavior analysis.

## Conclusions

At the vanguard of the science gateway community, nanoHUB has established itself as a leader both in hosting of scientific code and development of a production-level, open source cyberinfrastructure platform. The nanoHUB team continuously considers and adapts to the rapidly evolving challenges facing the scientific and software community. We have found that a flexible approach safeguarding the rights and concerns of software authors while striving to quickly bring quality codes to a larger audience leads to the best potential for accelerating the transition of research from the labs and the scientists to the classroom and the greater scientific community.